\documentclass[twocolumn]{revtex4-1}
\usepackage{amsmath}
\usepackage{stmaryrd}
\usepackage{txfonts}
\usepackage{amssymb}
\usepackage{mathrsfs}
\usepackage{graphicx}
\usepackage{dcolumn}
\usepackage{bm}
\usepackage{braket}
\usepackage{color}
\usepackage{ulem}

\usepackage[colorlinks=true, linkcolor=blue, citecolor=blue, urlcolor=blue]{hyperref}
\usepackage[dvipsnames]{xcolor}
\begin{document}

\title{Reaching the equilibrium state of frustrated triangular Ising magnet Ca$_3$Co$_2$O$_6$}

\author{Ivan Nekrashevich$^{1,*}$}
\author{Xiaxin Ding$^{2,\&}$}
\author{Fedor Balakirev$^2$}
\author{Hee Taek Yi$^{3}$}
\author{Sang-Wook Cheong$^3$}
\author{Leonardo Civale$^1$}
\author{Yoshitomo Kamiya$^{4,\dagger}$}
\author{Vivien S. Zapf$^{2,\dagger}$}
\affiliation{$^1$Material Physics and Applications-Quantum (MPA-Q), Los Alamos National Laboratory, Los Alamos, NM, 87545, USA}
\affiliation{$^2$National High Magnetic Field Lab, Los Alamos National Laboratory, Los Alamos, NM, 87545, USA}
\affiliation{$^3$Rutgers Center for Emergent Materials and Department of Physics and Astronomy, Rutgers University, Piscataway, NJ 08854, USA}
\affiliation{$^4$School of Physics and Astronomy, Shanghai Jiao Tong University, Shanghai, China}
\affiliation{$^*$Now at Fermi National Accelerator Laboratory}
\affiliation{$^\&$Now at City College New York}
\footnote{$^\dagger$Co-last authors}

\begin{abstract}
Ca$_3$Co$_2$O$_6$ is a frustrated magnet consisting of a triangular arrangement of chains of Ising spins. It shows regular magnetization steps vs magnetic field every 1.2 T that are metastable with very slow dynamics. This has puzzled the community for many years and given rise to numerous potential theories. Here we approach the problem by seeking the elusive magnetic equilibrium state at $T =$ 2 K. To this end, we explore two approaches: (1) bypassing the slow dynamics produced by changing fields by instead field-cooling directly to the target temperature, and (2) quantum annealing in transverse magnetic fields. 
While we observe no measurable effect of the quantum annealing in fields up to 7 T, which is likely due to the large Ising anisotropy of Co spins in this material, we find that for the field cooling in longitudinal fields we achieve the predicted equilibrium 1/3 magnetization. We perform Monte Carlo simulations of the ground state phase diagram and we also simulate the quantum annealing process and find good agreement between experiment and theory. Thus we present an investigation of the elusive ground state properties of the canonical frustrated triangular system Ca$_3$Co$_2$O$_6$.

\end{abstract}
\date{\today}
\maketitle

\section{Introduction}

In Ising frustrated magnets, seemingly simple frustrated magnetic interactions can give rise to complex magnetic behaviors including large unit cells, incommensuration, fractal phase diagrams, spin liquid behavior, and very slow dynamics~\cite{Wannier50,Bak82,Selke88,Coppersmith85,Ramirez94,Balents10}. Ca$_3$Co$_2$O$_6$ is an example of frustrated behavior in Ising spins with slow dynamics that has remained a puzzle for 25 years~\cite{Fjellvag96,Aasland97,Kageyama97}. The structure consists of ferromagnetically-coupled chains of Ising spins along the $c$ axis with alternating prismatic Co$^{3+}$ $S = 2$ sites and octahedral $S = 0$ Co$^{3+}$ sites, and these chains are arranged in a triangular configuration in the $ab$ plane with antiferromagnetic inter-chain coupling~\cite{Fjellvag96}.  
The magnetization vs magnetic field, $M(H)$, shows regular steps every 1.2 T up to 3.6 T, followed by less distinct, irregular steps leading to the saturation field of 7 T~\cite{Hardy04}. However, these steps do not correspond to states in equilibrium. At 4 K and above, magnetic fields applied at extraordinarily slow sweep rates of 0.01 T/min cause the step phase to disappear in favor of a single 1/3 plateau in $M(H)$ followed by a step to saturation~\cite{Hardy04}.  This 1/3 plateau state is naturally expected to be a ground state of a frustrated triangular Ising system in a magnetic field \cite{Qin09}, in which two spins in a triangle point up and one points down. The sweep rate needed to induce the metastable step phase varies with temperature, extending beyond experimentally feasible time scale in temperatures below 4 K, such that the equilibrium state below 4 K has never been measured.  

In zero field, long-wavelength incommensuration along the $c$-axis chains and short-range order has also been observed by neutron and X-ray diffraction measurements
below $T_c \sim$ 25K ~\cite{Agrestini08a,Agrestini08b,Mazzoli09,Fleck10,Agrestini11,Moyoshi11,Motoya18}. The incommensurate wavevector slides towards simple ferromagnetism along the chains as the temperature is lowered, and also evolves with time over several days. 

Many theories have been proposed to explain the peculiar magnetic behavior of Ca$_3$Co$_2$O$_6$. Some theories predicted a partially-disordered antiferromagnetic state~\cite{Kageyama97}, where each ferromagnetic chain along the $c$-axis is treated as a giant Ising spin that forms different patterns of up and down in the ab plane \cite{Yao06,Kudasov06,Soto09}.  
A different theory is inspired by single-molecule magnets where the steps in the magnetization result from quantum tunneling \cite{Chapon09}. These theories all predate the observation of incommensuration in the $c$-axis chains, which later on prompted Kamiya (co-author of this work) and Batista~\cite{Kamiya12} to propose a modified version of the axial next-nearest-neighbor Ising (ANNNI) model. In the ANNNI model, frustration occurs between ferromagnetic and antiferromagnetic nearest and next-nearest neighbor interactions $J_\mathrm{nn}$ and $J_\mathrm{nnn}$ along the chains. This gives rise to a fractal phase diagram with a theoretically infinite number of phases as a function of temperature $T$ and exchange interactions with different incommensurate wave vectors~\cite{Bak82,Selke88}  (in other words, frictional sliding of the incommensurate wave vector). As the temperature is lowered, the wave vector approaches commensuration, while at the lowest temperatures a commensurate ground state with 'up up down' or 'up up down down' spins appears depending on the ratio of $J_\mathrm{nn}/J_\mathrm{nnn}$. In Ref.~\onlinecite{Kamiya12}, it was argued that an effective mean-field description very similar to that of the classic ANNNI model can be derived for Ca$_3$Co$_2$O$_6$ by interleaving three chains to form one effective chain with frustrated nearest, next-nearest and next-next-nearest neighbor interactions. Quantum Monte Carlo (MC) simulations performed in the original lattice for Ca$_3$Co$_2$O$_6$ indeed reproduced both the long-wavelength SDW order and the regular magnetization steps below 3.6 T~\cite{Kamiya12}. 

However, to bridge a missing link between theory and experiments, a new experimental protocol has to be developed to reach the low-$T$ equilibrium state (minimum of the free energy). In this work, we present the results of our investigation of two methods to reach the ground state: the first method is to cool the system in a longitudinal field at each magnetic field, thereby avoiding the slow dynamics that occurs when the field is changed at low temperatures. The second method is to quantum anneal the system with magnetic fields transverse to the Ising axis~\cite{Brooke99,King21}, applied either while cooling or after cooling to 2 K. We compare our experimental data with our MC simulations for this system.

\begin{figure}[htbp]
\includegraphics[width=0.9\hsize]{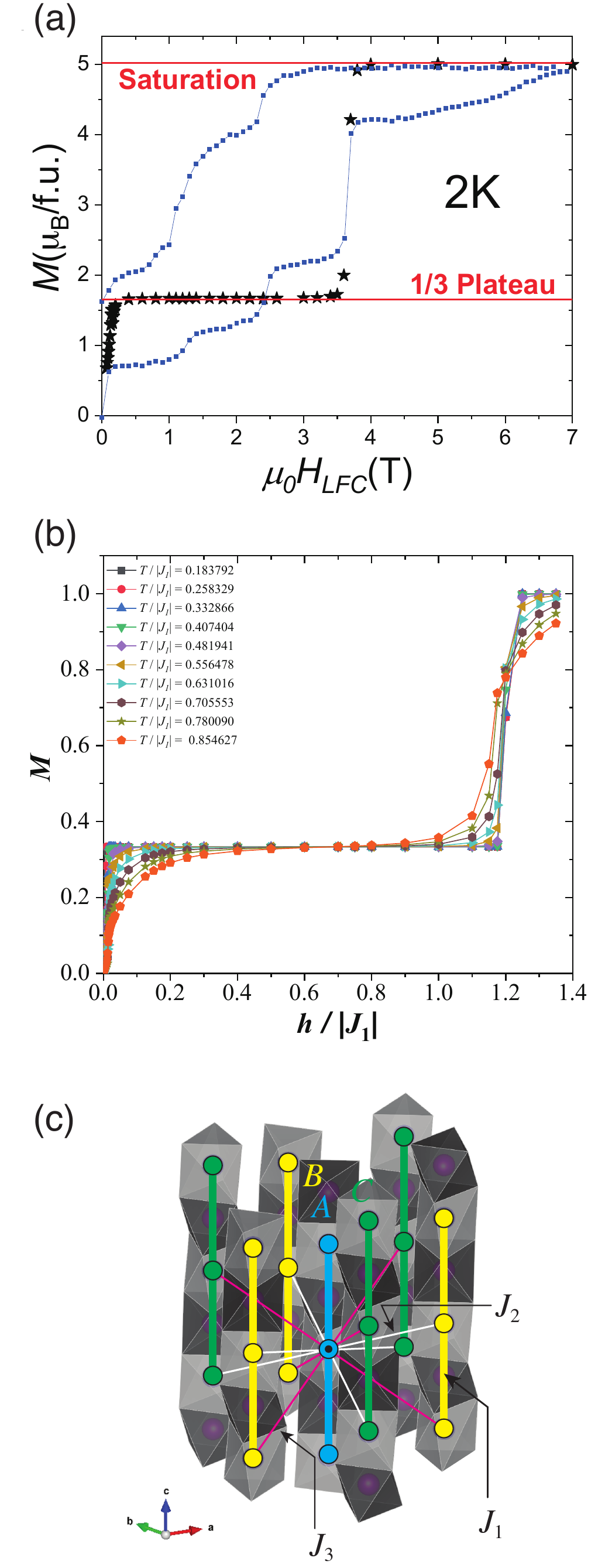}
\caption {(color online) 
(a) Experimental Magnetization $M$. Black stars (ground state): $M$ after cooling from 30 to 2 K in a longitudinal field ($H_{LFC}$) $H\parallel c$ and then measuring at the same field. Each star represents a separate field cool. Blue squares (metastable): $M(H)$ measured by cooling in zero field from 30 to 2 K and then sweeping the field to 7 T and back to 0.
(b) Ground state $M(H)$ obtained by classical MC simulations.
(c) Schematic picture of the magnetic lattice of Ca$_3$Co$_2$O$_6$, with the intrachain coupling $J_1$ and the interchain couplings $J_2$ and $J_3$, superimposed on the crystal structure. For the sake of clarity, only $J_2$ and $J_3$ connected to the site in the center are shown.
}
\label{plateau}
\end{figure}

\section{Experimental details}
Magnetization was measured in a 7 T Magnetic Property Measurement System (MPMS) Superconducting Quantum Interference Device (SQUID) by Quantum Design with a horizontal-axis sample rotator. Magnetostriction measurements were carried out using a piezoelectric strain gauge~\cite{PSG}  in a 14 T Physical Properties Measurement System by Quantum Design.
The sample with a crystalline face was mounted on the sample holder with less than half a degree of misalignment. For magnetization measurements, additional alignment to correct for the thermal contraction of the rotator was achieved for longitudinal fields by measuring the sample's remanent moment after field cooling as a function of angle and seeking the maximum of the projection on the vertical axis. For quantum annealing experiments, aligning the c-axis precisely perpendicular to the field was more tricky and a greater degree of alignment is needed to avoid accidental longitudinal fields. The procedure is described in the quantum annealing section below. 

\section{Equilibrium magnetization curve at low temperature}
\subsection{Experiments}
In Fig.~\ref{plateau}(a), we show magnetization data vs magnetic field. The star symbols (black) show the magnetization measured after cooling from 30 K to 2 K in a longitudinal field ($H_{LFC}$) to the final measurement temperature. In other words, each data point is measured after separate field cooling at a different magnetic field to avoid sweeping the field at low temperatures. The data shows a plateau at 1/3 of the saturation magnetization, 1.66 $\mu_B$, before saturating above 3.6 T.
This data is in stark contrast with the previously-published $M(H)$ curves measured by sweeping the field after zero-field cooling (ZFC), 
as also reproduced here with square symbols (blue). As is well-known, the field-sweep data shows a series of steps every 1.2 T due to the very slow dynamics that traps the system into metastable states.\cite{Hardy04} At 4 K, Hardy \textit{et al}.~showed that the slow dynamics can be overcome and the 1/3 plateau state can be reached by sweeping the field very slowly at less than 0.01 T/min~\cite{Hardy04}. However, the 1/3 plateau thus observed is not as flat as the one at 10~K, implying that slow dynamics could still be preventing the system from showing true thermal equilibrium. Below 4 K, the dynamics diverge such that a week-long field sweep still shows the steps. Our direct field-cooling protocol, on the other hand, can bypass the slow dynamics and bring the system to its equilibrium even at temperatures as low as 2 K.

\subsection{MC simulations}
In Fig.~\ref{plateau}(b), we show magnetization obtained by classical MC simulations for this system at low temperatures. Here we consider the same Hamiltonian that was previously studied~\cite{Kamiya12}, defined in the same three-dimensional lattice as Ca$_3$Co$_2$O$_6$, though we include no transverse field term in this study.
Trigonally prismatic Co$^{3+}$ ($3d^6$) sites with $S = 2$ has large easy-axis anisotropy, which permits a description with an effective Ising model,
\begin{align}
    \hat{H} = \sum_{\gamma = 1,2,3}\sum_{\langle{ij}\rangle_{\gamma}} J_\gamma \sigma^z_i \sigma^z_j - h \sum_i \sigma^z_i,
    \label{eq:H}
\end{align}
where $\sigma^z_{i} = \pm 1$ represents the ground state doublet at site $i$, $\langle{ij}\rangle_{\gamma}$, $\gamma \in \{1,2,3\}$ stands for neighboring sites connected by $J_\gamma$, and $h = g \mu_B S H$ is the magnetic field.
$J_1 < 0$ is the ferromagnetic intrachain interaction, whereas $J_2$ ($J_3$) is the antiferromagnetic interchain interactions with the vertical shift of 1/3 (2/3) lattice units along the $c$ axis, as shown in Fig.~\ref{plateau}(c). To account for the highly anisotropic nature of the spin correlation in this material, we consider a lattice of size $L \times L \times L_c$ with $L = 4$ and $L_c = 40L$ ($N_\text{spin} = 3L^2 L_c = 7680$ spins) with periodic boundary conditions.
We set $J_2 = J_3 = 0.1\lvert{J_1}\rvert$ to obtain the data in Fig.~\ref{plateau}(b), which is motivated by the following considerations. Firstly, $J_2 \simeq J_3$ is suggested by an \textit{ab initio} study~\cite{Fresard04}. Secondly, an NMR experiment reported $J_1 = -23.9(2)$\,K and $J_2 + J_3 = 2.3(2)$\,K, further suggesting $J_2 = 1.1$\,K and $J_3 = 1.2$\,K to explain the ordering vector~\cite{Allodi14}. Here, although the ratio $J_2 / \lvert{J_1} \approx J_3 / \lvert{J_1} \approx 0.046$ suggested in this way is even smaller than the one considered in this work, it has been confirmed that the model can reproduce the same kind of long wavelength three-sublattice SDW order (wavelength $\sim$ 80 sites) below $T_c / \lvert{J_1}\rvert \approx 1.4$, as will be discussed elsewhere. The data shown in Fig.~\ref{plateau}(b) was obtained by combining single-spin and intra-chain cluster updates~\cite{Kamiya12} as well as replica exchanges~\cite{Hukushima96} performed every 10 MC steps. A few hundred replicas and careful tuning of their individual temperatures are needed for this system size to address equilibrium properties at low temperatures. The temperature dependence of the raw data thereby obtained is smooth enough at each field to allow us to perform interpolation to evaluate $M(H,T)$ at several (equally separated) temperatures shown in Fig.~\ref{plateau}(b). In this way, we confirm that the system at low enough temperatures realizes the 1/3 magnetization over a wide field range. Here, the lowest $T$ is $T / \lvert{J_1}\rvert = 0.183792$, which amounts to $T / T_c \approx 0.13$. By taking into account the transition temperature $T_c \approx$ 25 K in this material, the lowest $T$ roughly corresponds to $\approx$ 3.3~K.

\begin{figure}[htbp]
\includegraphics[width=0.5\textwidth]{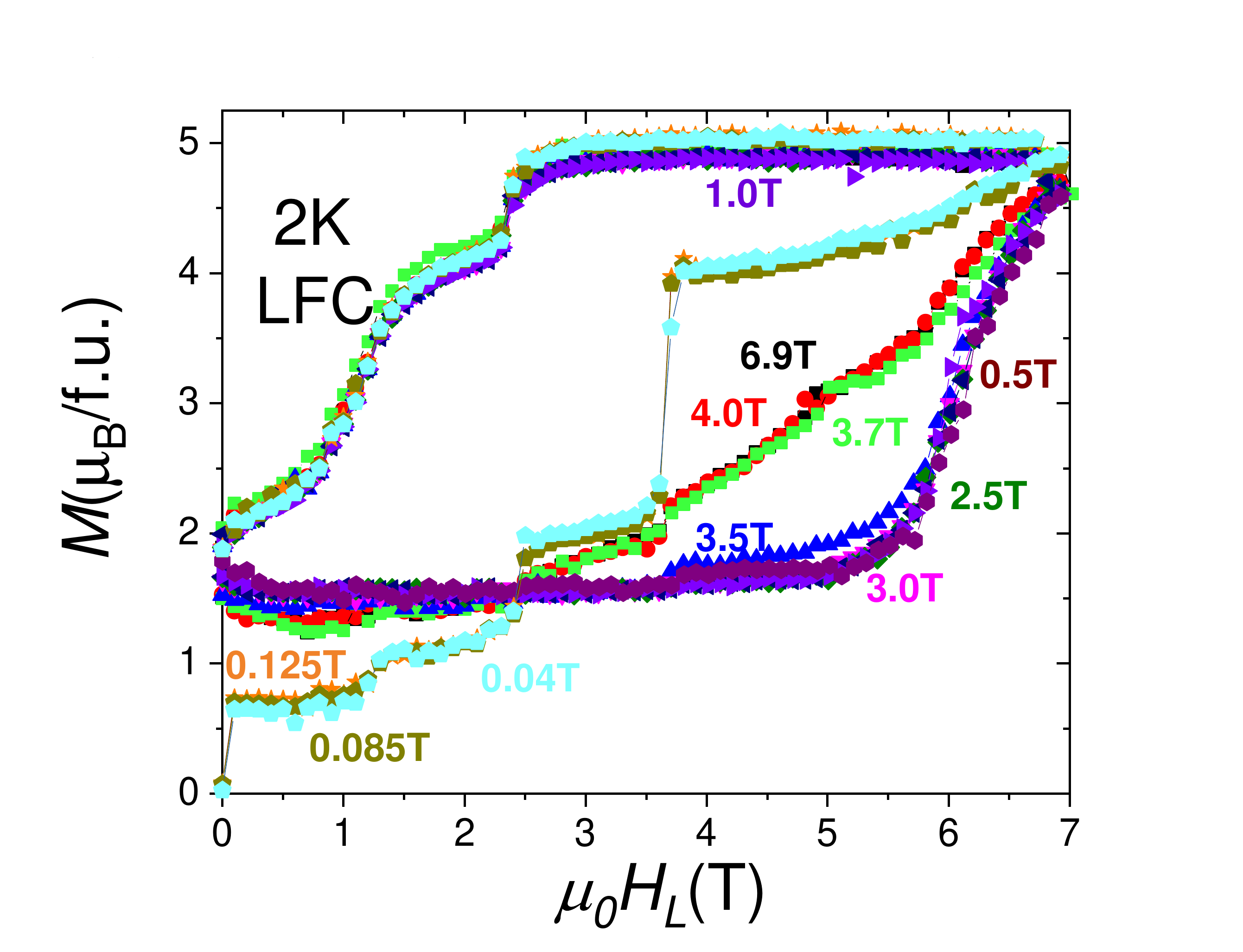}
\caption {(color online) $M(H)$ measured while sweeping the longitudinal field, after cooling from 30 to 2 K in different longitudinal fields as labelled in the plot. Before measuring each M(H) loop, the FC field is switched off. The magnetic field sweep rate was 10 Oe/s.}
\label{loops}
\end{figure}

\section{History necessary to reach the metastable step state or the 1/3 plateau state}
In this section we explore which history of field and temperature evolution is needed to achieve the equilibrium 1/3 plateau vs the metastable step state. In particular, we find that certain protocols yield a stable 1/3 plateau that persists even after subsequent field sweeps. 
In Fig.~\ref{loops}, we first cool the system in different longitudinal fields ($H || c$) from 30 K to 2 K, turn off the field, and subsequently sweep the longitudinal $H||c$ magnetic field from 0 to 7 T at 10 Oe/s. For the initial magnetic field between 0 and 0.125 T applied during the cooling stage, the step phases are observed on subsequent field sweeps. However, when cooling in fields between 0.15 and 3.6 T (i.e., inside the region for the 1/3 plateau of Fig.~\ref{plateau}), we observe the 1/3 plateau state. It should be noted that here the step to saturation occurs at higher fields than the equilibrium data in Fig.~\ref{plateau}. This implies that the step to saturation is also delayed by slow dynamics, as discussed recently~\cite{Hegde20}.

\begin{figure*}[htbp]
    \begin{center}
    \includegraphics[width=\hsize]{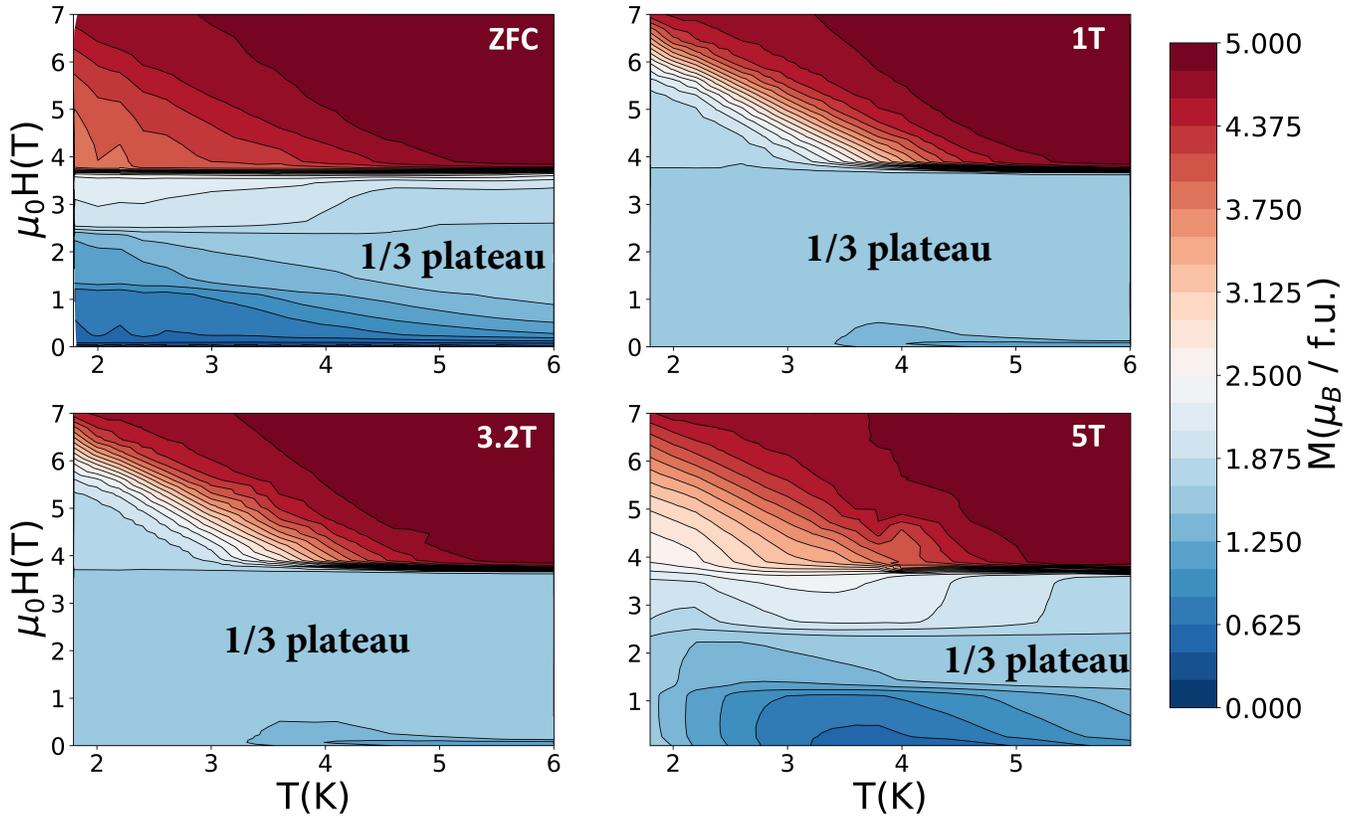}
\end{center}
\caption{(color online) $M(H)$ after field cooling in $H\parallel c$, showing the conditions necessary to observe the 1/3 plateau. After field-cooling to the target temperature, the magnetic field was switched off and subsequently, longitudinal $H || c$ magnetization vs field was measured starting from 0 to 7 T. Upsweep date is shown. The color is the value of $M$. The magnetic fields applied during the field-cooling step are shown in the top right corners of the diagrams. Vertical axes show the swept longitudinal magnetic field of the M(H) curves. The horizontal axis shows the target temperature at which M(H) upsweep was performed after field cooling. The value of the magnetization at 1/3 saturation is light blue and labelled "1/3 plateau".}
\label{3D}
\end{figure*}

In Fig.~\ref{3D} we show the same data over a more extended region of fields and temperatures in a 3D color plot. E.g. after field cooling to the target temperature, the magnetic field was switched off and then the longitudinal field was swept from 0 to  7 T. The target temperatures ranged from 1.8 K to 6.0 K and field-cooling fields of $H=$ 0, 1 T, 3.2 T and 5 T were used. In Fig.~\ref{3D} the color is the magnetization, the y-axis shows the swept field, the temperature is the target temperature, and the inset shows the field-cooling field.

Our experiments show that a 1/3 plateau is achievable for fields swept at low temperatures, if initially a high enough magnetic field $H\parallel c$ is applied during the field cooling. Also, it can be noted that the onset of the transition to saturation does not depend on either the magnitude of the field during the cool down nor on the final temperature, and appears at ~3.6 T during the upsweep $M(H)$ measurement. Applying field higher than 3.6 T during the initial field cooling brings the system in the saturated state and the subsequent M-H upsweeps at different temperatures show re-emergence of the step phase. For 5 T FC, we observe the 1/3 plateau at high temperatures, as is consistent with previous works \cite{Hardy04}. However we were also surprised to observe a narrow 1/3 plateau below 2 K and the steps phase is the most pronounced in the region of temperatures 3 - 4 K.

\begin{figure}[htbp]
\includegraphics[width=0.5\textwidth]{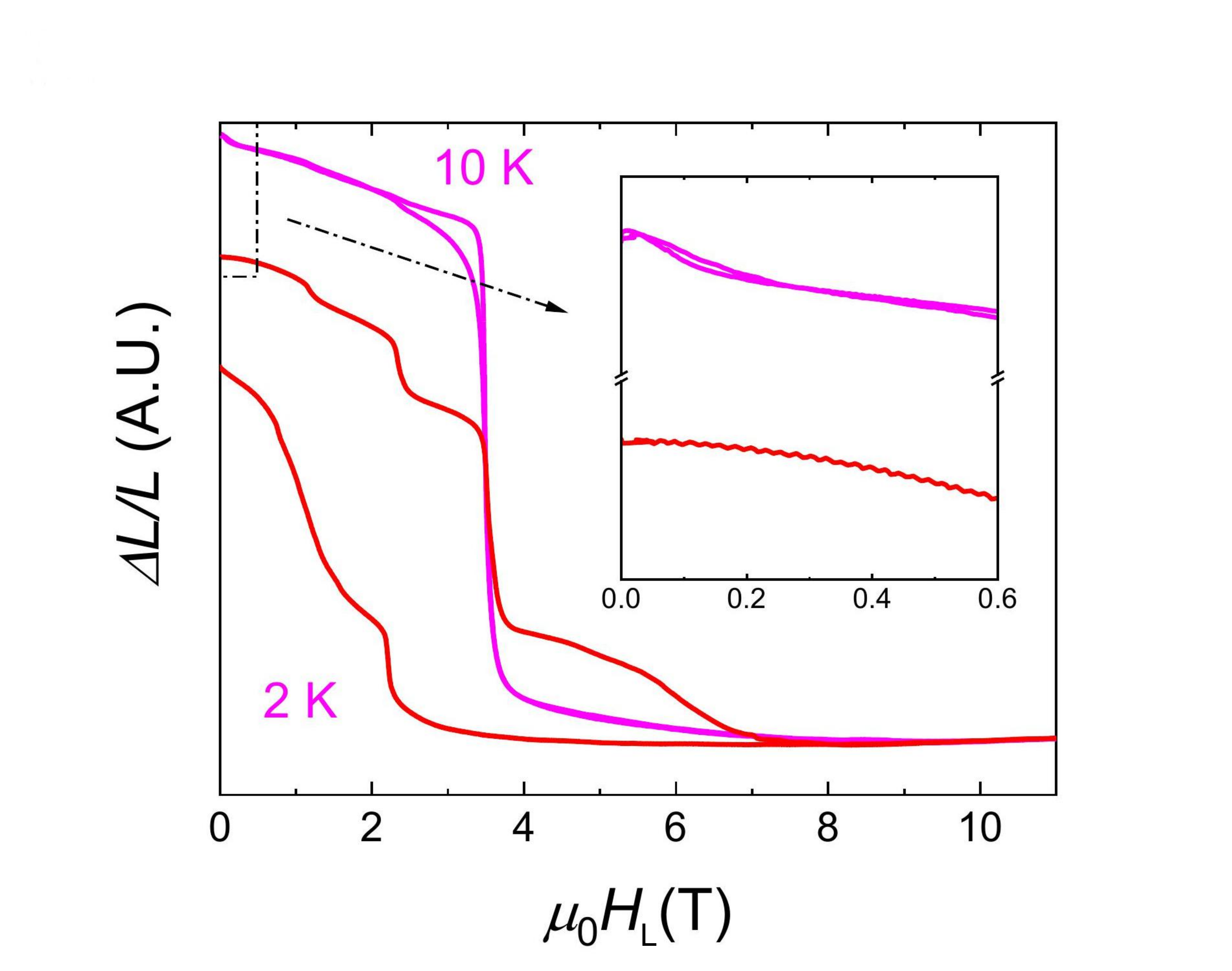}
\caption {(color online) Field dependence of magnetostriction at 2 and 10 K with longitudinal magnetic field $H_L$ along $c$-axis and swept at 100 Oe/s. Inset is the amplified view of low-field data showing that the initial step that was seen in the $M(H)$ data is mostly absent in the $\Delta L/L$ data.}
\label{magnetostriction}
\end{figure}

In addition to magnetization we have also measured the magnetostriction $\Delta L/L$ of Ca$_3$Co$_2$O$_6$ with $H || c$ after zero field cooling at different temperatures, as plotted in Fig.~\ref{magnetostriction}. Magnetostriction in insulating magnets reflects the change in length of the sample as a result of changes in bond lengths, which occur to minimize the magnetic energy via changes in the exchange interaction and anisotropies.  At 10 K, the magnetostriction shows all the metastable steps seen in the $M(H)$ curve. However in the inset of Fig.~\ref{magnetostriction}, the first jump ($\approx$0.15 T) in $\Delta L/L$ at 10 K is very small while there is no jump at 2 K. The discrepancy between the $M(H)$ data and $\Delta L(H)$ data is not due to the change in field sweep rate (10 Oe/s vs 100 Oe/s) because the initial step is also seen at 100 Oe/s in previously published magnetization data \cite{Hardy04,Zapf18}. One possible reason is that the first step is due to domain alignment rather than due to a change in the microscopic magnetic order. Magnetostriction is insensitive to domain alignment since only the domain boundaries are affected and they occupy a very small percentage of the sample's volume. For example, the predicted equilibrium 1/3 state in the magnetization has a net moment and thus could form domains. T

\begin{figure}[htbp]
\includegraphics[width=0.5\textwidth]{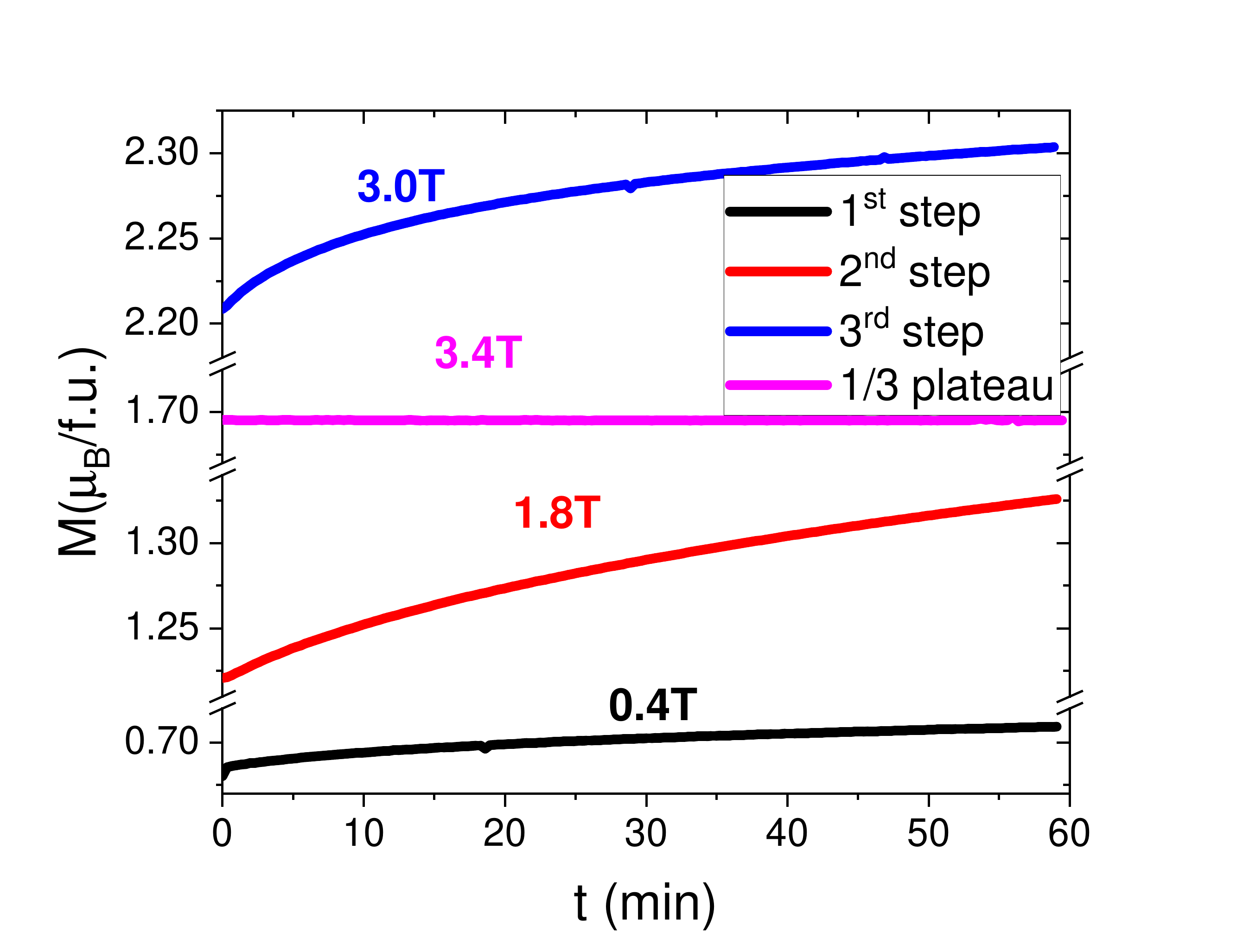}
\caption {(color online) Time-relaxation curves for four magnetic states of  the system. Magnetization steps 1 - 3 (see Fig.~\ref{plateau}) are reached by cooling the sample in zero field from 30K to 2K and then ramping $H || c$  to 0.4 T, 1.8 T and 3.0 T respectively at the rate 10 Oe/s. Subsequently, $M(H)$ collected as a function of time for one hour. The 1/3 plateau is reached by cooling the sample from 30K to 2K in 3.4 T parallel to the c-axis. The vertical axis has 3 breaks to ease comparison of the relaxation curves, and the vertical scale is constant throughout.}
\label{TR}
\end{figure}

In order to further investigate whether the 1/3 plateau is the equilibrium ground state, we performed magnetic relaxation experiments in the 1/3 state and in the metastable step state. Some of these have also been reported previously ~\cite{Moyoshi11}. After reaching each of the 3 steps, the system's magnetization was observed for 1 hour while temperature and magnetic field were kept constant (black, red and blue curves on Fig.~\ref {TR}). At all three steps, slow evolution of the magnetic moment is evident. In the case of the 1/3 plateau, however, the magnetization remained stable during the 1 hour measurement. Such a high stability of the magnetization is indicative of equilibrium achieved by the spin system and suggests that 1/3 plateau is the global energy minimum state. A more detailed time-evolution of the CCO spin system will be provided by us elsewhere.

\section{Quantum Annealing}
Besides exploring longitudinal field cooling to reach the equilibrium state, we have also explored transverse field cooling to see if we can observe quantum annealing. We are motivated in part by simulations of this compound using a D-WAVE quantum annealing computer, which found the 1/3 equilibrium state shown in Fig.~\ref{plateau} \cite{King21}.

\begin{figure}[htbp]
\includegraphics[width=0.5\textwidth]{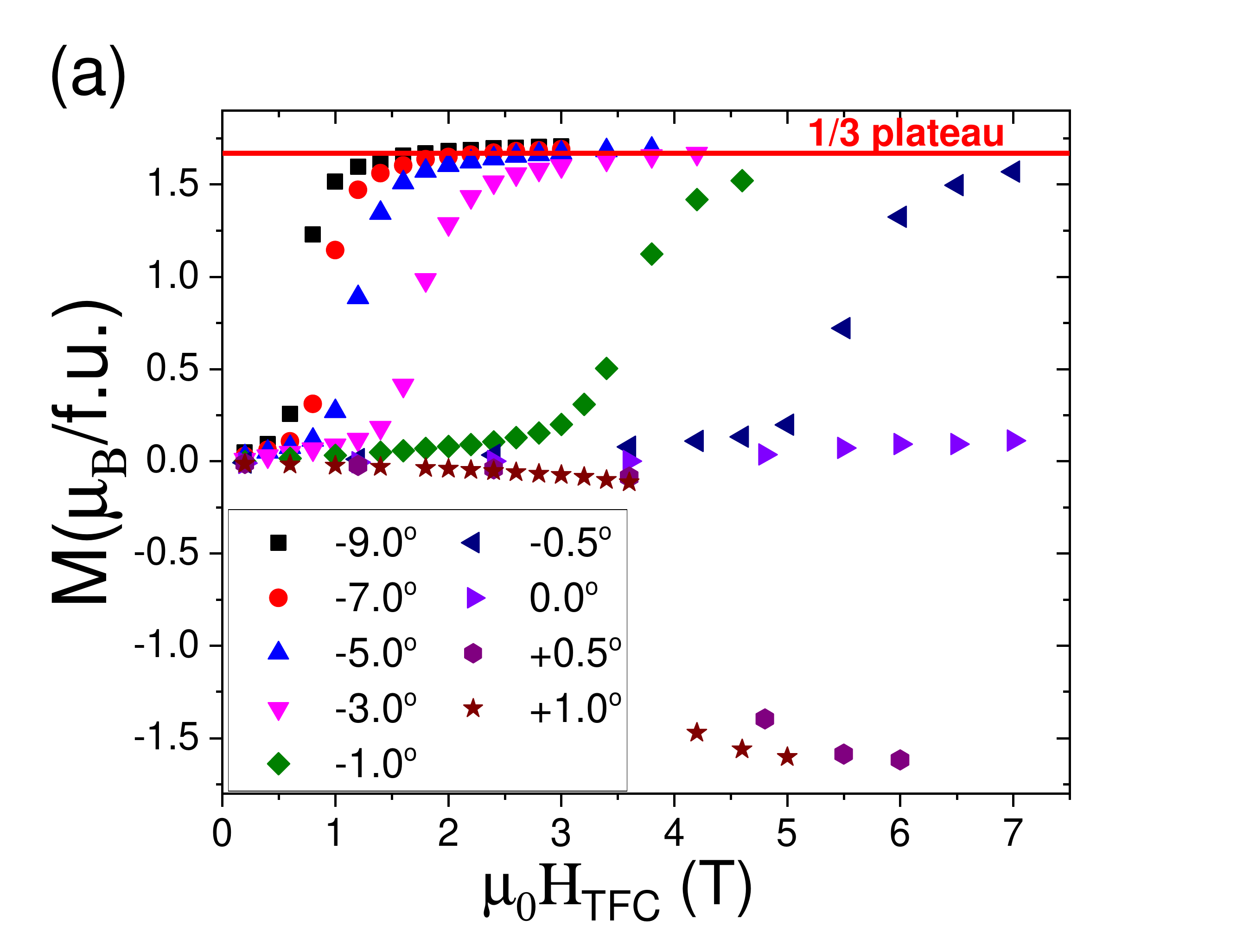}
\includegraphics[width=0.5\textwidth]{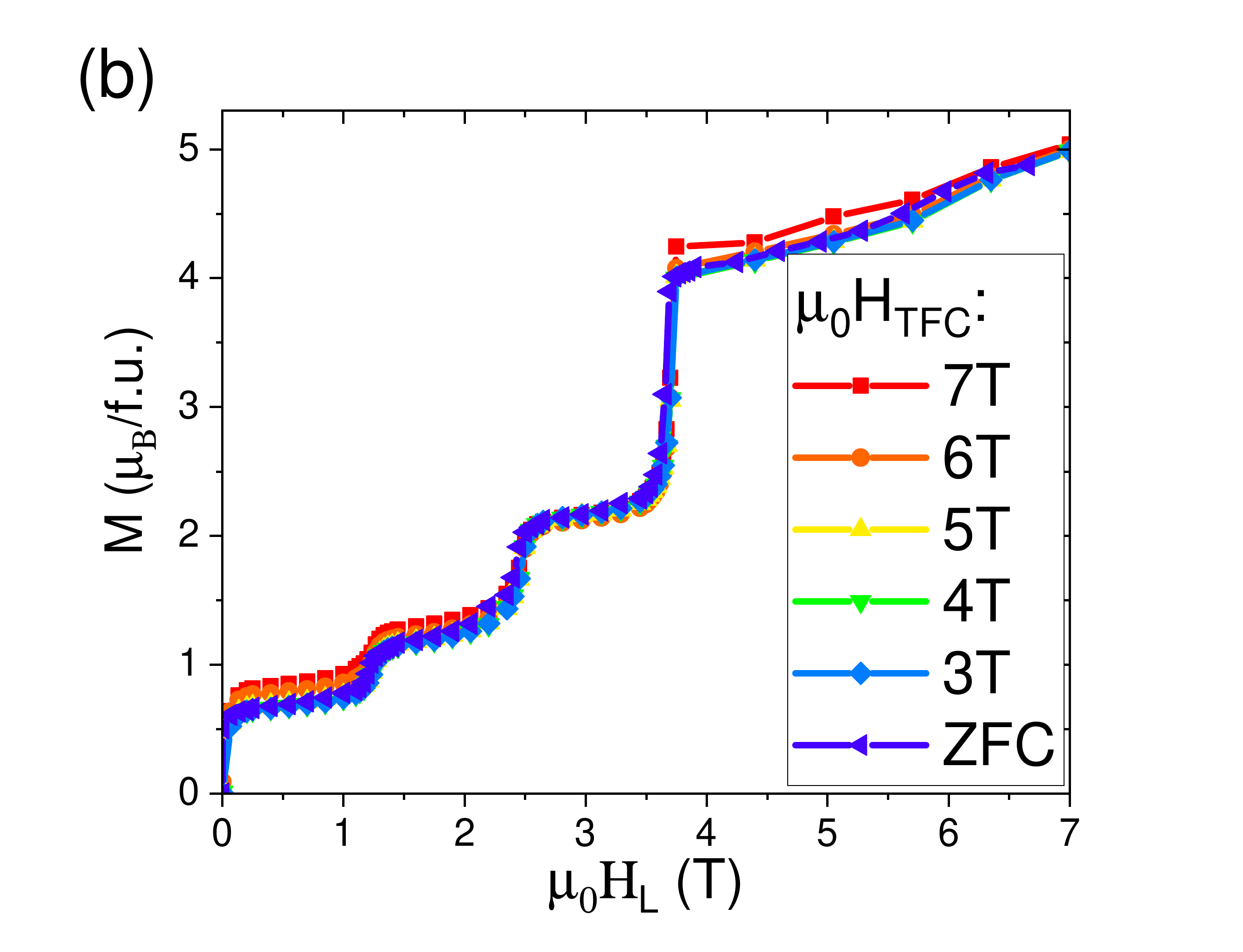}
\caption {(color online) (a) Remanent magnetization measured after cooling from 30K to 2K in various fields applied at orientations close to ab-plane (from nominally -9.0{\textdegree} to +1.0{\textdegree}). Each data point represents the remanent magnetization after cooling the sample (magnitude of the magnetic field during the cool-down step is shown on horizontal axis) in a mostly transverse field at various orientations (groups of points of different color) relative to the ab-plane of the CCO crystal. The change of the sign of remanent magnetization is due to the change of the projection of magnetic field on the c-axis during the cool-down. The orientation of the field that didn't yield transition to 1/3 plateau even at 7 T was assumed {$H\perp$c}. This orientation marked by the curve obtained at 0.0{\textdegree} (nominal angle between H and ab-plane).
Figure (b) shows $M(H)$ loops obtained by longitudinal field sweep after cooling from 30K to 2K in the transverse field identified in (a), showing no evidence of quantum annealing.}
\label{TFC}
\end{figure}

Here we explore quantum annealing on actual crystals of Ca$_3$Co$_2$O$_6$. However we do not see definitive evidence of quantum annealing for fields up to 7 T. We attempted quantum annealing by two methods: in the first we zero-field-cooled the sample to 2 K and then applied a transverse magnetic field $H_{\mathrm{TFC}}$, $H \perp c$. In the second method, we cooled the sample in $H_{\mathrm{TFC}}$ from 30 to 2 K. For this second method, the alignment of the sample in the field is critical as any misalignment will create longitudinal field cooling. To achieve optimal alignment we systematically attempted transverse field cooling different alignments close to $H \perp c$. We found that there was one orientation that produced no remanent magnetization and produced no steps in the subsequent $H || c$ $M(H)$ data. We took that field to be the transverse field. In Fig.~\ref{TFC}a, we show the remanent moment $M_r(H_{\mathrm{TFC}})$ data taken after field cooling in different transverse fields $H_{\mathrm{TFC}}$ and for different nominal alignments. The optimal angle is is purple triangles labeled 0.0$^{\circ}$ angle.

After field cooling in this optimal transverse orientation we measured $M(H)$ in longitudinal fields to see whether quantum annealing was effective at producing the 1/3 plateau state, or if the sample would show the metastable step state. In  Fig.~\ref{TFC} (b) we show that the metastable step state resulted, i.e. showing no evidence of quantum annealing towards the equilibrium state.  Fig.~\ref{TFC} (b) shows $M(H)$ upsweeps with longitudinal field ($H||c$) after CCO crystal was cooled from 30K to 2K in the optimal transverse fields of either 0, 3, 4, 5, 6, or 7 T. Cooling in transverse fields up to 7 T produced no effect on the magnetization curves - the metastable step state is observed for all transverse field coolings.

\begin{figure}[t]
\includegraphics[width=\hsize,bb=0 0 324 252]{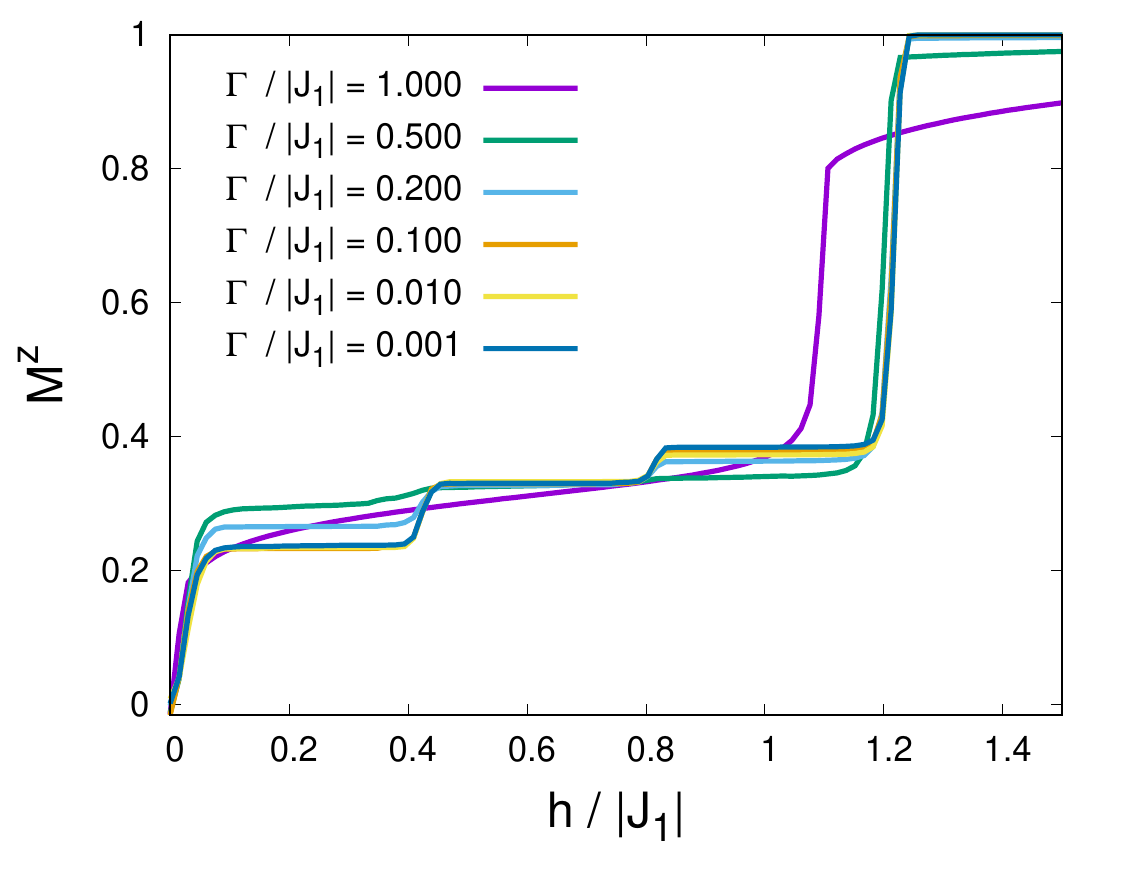}
\caption {(color online) 
Out-of-equilibrium magnetization computed by quantum MC simulations for different values of $\Gamma$ and $J_2 = J_3 = 0.1 \lvert{J_1}\rvert$. The system size is 12$\times$12$\times$120 (3 sites per unit cell). The system is initially thermalized at $h = 0$ and $T / \lvert{J_1}\rvert = 0.3$ before performing an up-sweep of $h$.}
\label{fig:QA}
\end{figure}

We have performed quantum MC simulations by including a transverse field term $\hat{H}_\Gamma = -\Gamma \sum_i \sigma^x_i$ in Eq.~\eqref{eq:H}. The system size in this simulation is 12$\times$12$\times$120 and the coupling constants are the same as above ($J_2 = J_3 = 0.1 \lvert{J_1}\rvert$). The protocol for this numerical experiment follows one in Ref.~\onlinecite{Kamiya12}: we first thermalize the system at $h = 0$ and $T / \lvert{J_1}\rvert = 0.3$ and then perform a up-sweep of $h$. This indeed results in out-of-equilibrium magnetization steps that becomes smaller for larger values of $\Gamma$ (Fig.~\ref{fig:QA}). However, we find that the typical energy scale to achieve a sizable quantum annealing effect is $\Gamma / \lvert{J_1}\rvert \gtrsim 0.5$ or more. In experiments however, the size of $\Gamma = g\mu_B\mathbf{S}\cdot \mathbf{B}$ for a 7 T transverse field is 300 times smaller than $J_1$. This is because the transverse spin $\mathbf{S}$ is less than 1\% of the longitudinal spin as measured in our samples. Thus the transverse field needed to achieve quantum annealing in experiments would be at least 2,100 T.

Quantum annealing in a model version of Ca$_3$Co$_2$O$_6$ was also explored by King et al \cite{King21} using both quantum MC simulations and simulations on a 2D D-wave quantum annealing computer. However there also the size of the transverse field is much larger than in our experiments. Their value of $\Gamma$ is comparable to the energy scale of the longitudinal field needed to induce saturation, or about 100 times larger than in our experiments.

\section{Conclusion}
In conclusion, we have measured and simulated the magnetic properties of the triangular frustrated Ising spin chain system Ca$_3$Co$_2$O$_6$. We have addressed the long-standing puzzle of finding the experimental ground state at low temperatures by field cooling at each temperature. Thereby we avoid the extremely slow dynamics that so far have prevented reaching equilibrium when the field is swept at low temperatures. 
We have performed MC simulations that demonstrate good agreement with our experimental data showing 1/3 plateau as the equilibrium state of the system. We have explored and identified the range of the field cooling parameters such as longitudinal field amplitude and final temperature that bring the system into the 1/3 equilibrium magnetic state.  Finally we note that after an extensive and careful effort to achieve the necessary alignment, quantum annealing with a transverse magnetic field up to 7 T was not effective in reaching the equilibrium ground state, likely due to the very small magnetic susceptibility transverse to the Ising axis. This is in agreement with our quantum MC simulations of the quantum annealing process, as well as simulations in the literature \cite{King21} that require thousands of Tesla of transverse field to achieve quantum annealing.
Thus we have advanced our understanding of the equilibrium ground state of this triangular frustrated Ising system, which is an implementation of the classic ANNNI model for frustrated spins.

\begin{acknowledgments}
We are grateful to Cristian Batista for valuable discussions. V.S.Z. and X. D. were funded by the LDRD program at LANL. The facilities of the NHMFL are funded by the U.S. NSF, the DOE and the State of Florida through Cooperative Grant No. DMR-1157490. I.N. and L.C. were supported by the US D.O.E, Office of Basic Energy Sciences, Division of Materials Sciences and Engineering. Single crystal growth effort at Rutgers was supported by the DOE under Grant No. DOE: DE-FG02-07ER46382.
Y.K.~acknowledges the support by the NSFC (No.~11950410507, No.~12074246, and No.~U2032213), NSFC U2032213 and MOST (No.~2016YFA0300500 and No.~2016YFA0300501) research programs.
\end{acknowledgments}
\newpage
\renewcommand{\thesection}{\Alph{section}}

\end{document}